\documentclass[conference]{IEEEtran}
\IEEEoverridecommandlockouts

\usepackage{cite}
\usepackage{amsmath,amssymb,amsfonts}
\usepackage{algorithmic}
\usepackage{graphicx}
\usepackage{textcomp}
\usepackage{xcolor}
\usepackage{multirow}
\usepackage{algorithmic}
\usepackage{algorithm}
\usepackage{subfig}
\PassOptionsToPackage{hyphens}{url}\usepackage{hyperref}
\def\BibTeX{{\rm B\kern-.05em{\sc i\kern-.025em b}\kern-.08em
    T\kern-.1667em\lower.7ex\hbox{E}\kern-.125emX}}
\begin{document}
\small
\title{Frequency-Based Federated Domain Generalization for Polyp Segmentation
\thanks{This work was supported by the following grants: NIH NCI R01-CA246704, R01-CA240639, U01-CA268808, NIH/NHLBI R01-HL171376, and NIH/NIDDK \#U01 DK127384-02S1. The source code is available at \url{https://github.com/NUBagciLab/ICASSP2025-FDGPolyp}.}
}

\author{\IEEEauthorblockN{Hongyi Pan, Debesh Jha, Koushik Biswas, Ulas Bagci}
\IEEEauthorblockA{\textit{Machine \& Hybrid Intelligence Lab, Department of Radiology} \\
\textit{Northwestern University}\\
Chicago, Illinois, USA \\
\{hongyi.pan, debesh.jha, koushik.biswas, ulas.bagci\}@northwestern.edu}
}

\maketitle

\begin{abstract}
Federated Learning (FL)  offers a powerful strategy for training machine learning models across decentralized datasets while maintaining data privacy, yet domain shifts among clients can degrade performance, particularly in medical imaging tasks like polyp segmentation.  This paper introduces a novel Frequency-Based Domain Generalization (FDG) framework, utilizing soft-thresholding and hard-thresholding in the Fourier domain to address these challenges. By applying soft-thresholding and hard-thresholding to Fourier coefficients, our method generates new images with reduced background noise and enhances the model's ability to generalize across diverse medical imaging domains. Extensive experiments demonstrate substantial improvements in segmentation accuracy and domain robustness over baseline methods. This innovation integrates frequency domain techniques into FL, presenting a resilient approach to overcoming domain variability in decentralized medical image analysis. 

\end{abstract}


\section{Introduction}
Polyp segmentation in medical imaging is crucial for the early detection and treatment of colorectal cancer~\cite{brandao2017fully,akbari2018polyp,fan2020pranet,banik2020polyp,zhang2020adaptive,wei2021shallow}. Accurate identification and delineation of polyps can significantly impact patient outcomes, but the task remains challenging due to variations in imaging conditions, equipment, and patient demographics. Traditional machine learning approaches for polyp segmentation often face difficulties when applied to data from different sources or domains, leading to reduced performance and generalization issues. Another difficulty is more common: constructing a large, diverse dataset is essential for robust and successful models,  yet gathering such data in a centralized manner is often infeasible due to privacy laws and ethical concerns.

Federated learning (FL) has emerged as a promising solution to address data privacy concerns by allowing collaborative model training across multiple institutions without sharing sensitive data~\cite{li2020review,zhang2021survey,guo2022auto,jiang2023fair,gholami2024digest,miao2024federated}. In FL, local models are trained on decentralized datasets, and their updates are aggregated to form a global model. This method preserves patient privacy and enables the utilization of diverse datasets, which is essential for developing robust models. However, FL alone does not guarantee optimal performance across diverse domains. Domain generalization techniques aim to improve a model's ability to perform well on new, unseen domains by learning domain-invariant features. Despite these advancements, achieving consistent performance across varying imaging conditions remains challenging.

In this paper, we propose a novel approach that combines federated learning with frequency-based domain generalization techniques to enhance polyp segmentation. By incorporating frequency domain analysis, we aim to capture and leverage important features that contribute to improved domain generalization. Our approach addresses both the privacy concerns and the domain variability issues, leading to more accurate and robust polyp segmentation models. This innovative method, to our knowledge, is the first to combine these techniques, making it not only a powerful solution for polyp segmentation but also has a high potential to apply to a wide range of medical segmentation tasks such as retina fundus~\cite{wang2020dofe}, pancreas~\cite{khosravan2019pan}, breast~\cite{kramer2012pet}, liver~\cite{gorade2024towards,jha2024su1499}, brain~\cite{labella2024brain}, and others~\cite{rauniyar2023federated} due to its generic nature.

In the following, we begin by outlining the limitations of existing methods in handling domain variability and privacy issues. Next, we describe our federated learning framework and the frequency-based techniques used to improve domain generalization. Finally, we present experimental results demonstrating the effectiveness of our approach in achieving superior segmentation performance across diverse datasets. Our work contributes to the advancement of privacy-preserving, high-performance polyp medical segmentation.

\section{Methodology}
\subsection{Problem Statement}
In machine learning, traditional centralized approaches rely on aggregating data from various sources into a single location for training. However, in many real-world applications, especially in healthcare, data is distributed across multiple institutions or devices, and sharing this data is restricted due to privacy, regulatory, and security concerns. Federated learning (FL) addresses this issue by allowing local training on decentralized data while only sharing model updates, not raw data, across participants.

A critical challenge in FL is the ability to generalize across diverse domains or distributions, as data from different institutions or devices may come from varied populations, imaging modalities, or acquisition conditions~\cite{rauniyar2023federated}. This domain shift can cause models to perform well in some domains but poorly in others~\cite{huang2023rethinking}. Domain generalization aims to enable a model to perform well across unseen domains without retraining on new domain-specific data~\cite{zhang2024domain}.

Federated Domain Generalization (FDG) seeks to extend the federated learning paradigm by developing models that are inherently robust to domain shifts across different clients. The goal is to create models that can generalize well to new, unseen domains, making them more reliable and effective in real-world applications where heterogeneity between data sources is significant. Addressing this problem is crucial for deploying federated learning in sensitive fields like healthcare, where robustness across diverse patient populations and imaging devices is essential for safe and reliable diagnostic performance.

 Consider a set of $K$ training datasets $\{\Omega_0, \Omega_1, \dots, \Omega_{K-1}\}$ with corresponding annotations (\textit{e.g.}, segmentation masks) $\{\Psi_0, \Psi_1, \dots, \Psi_{K-1}\}$, and a test dataset $\Omega_K$ from an unseen domain. The objective in generalization is to develop a model trained on $\{\Omega_0, \Omega_1, \dots, \Omega_{K-1}\}$ that can effectively generalize to the completely unseen test domain, achieving high performance. To reach this objective, domain generalization techniques create a continuous domain distribution using the training datasets $\{\Omega_0, \Omega_1, \dots, \Omega_{K-1}\}$.

\subsection{Fourier-Transform-Based Domain Generalization}
A Fourier-Transform-based domain generalization (FDG) framework~\cite{xu2021fourier} leverages the discrete Fourier Transform (DFT) to improve generalization across multiple domains in machine learning tasks. This framework typically enhances the model's ability to learn domain-invariant features, which is essential for generalizing to unseen data distributions. The key idea of the FDG is that the amplitude spectrum obtained by the DFT contains the low-level distribution such as brightness, while the phase spectrum from the DFT stands for the high-level semantics such as object details of the image. Hence, one may replace the amplitude components of one domain's Fourier-transformed image with that of another domain to create a new augmented image.

In detail, given a source image $\mathbf{x}\in\mathbb{R}^{C\times H\times W}$ where $H\times W$ are the image resolution and $C$ is the channel number, its DFT $\mathbf{X}=\mathcal{F}(\mathbf{x})$ is computed as~\cite{brigham1967fast}:
\begin{equation}\label{eq: dft}
    \mathbf{X}(c, u, v) = \sum_{h=0}^{H-1}\sum_{w=0}^{W-1} \mathbf{x}(c, h, w)\cdot e^{-j2\pi\left(\frac{hu}{H}+\frac{wv}{W}\right)}.
\end{equation}

Let $\mathbf{A}=\text{abs}(\mathbf{X})$ and $\mathbf{P}=\text{angle}(\mathbf{X})$ denote the amplitude and the phase spectra of the source image, respectively. Transferring the style of the source image into the target with the amplitude of $\tilde{\mathbf{A}}$ is achieved as:
\begin{equation}
    \hat{\mathbf{A}} = (1-\lambda)\cdot\mathbf{A}+\lambda\cdot\tilde{\mathbf{A}},\label{eq: fda}
\end{equation}
\begin{equation}
    \hat{\mathbf{X}}(c, u, v) =  \hat{\mathbf{A}}(c, u, v)\cdot e^{-j\mathbf{P}(c, u, v)},
\end{equation}
where $\lambda$ controls the augmentation strength. In practice, $\lambda$ is dynamically set in the interval of $(0, 1]$. $\hat{\mathbf{X}}=\mathcal{F}(\hat{\mathbf{x}})$ stands for the DFT of the generated image $\hat{\mathbf{x}}$. Therefore, by taking the inverse discrete Fourier Transform the generated image is obtained:
\begin{equation}
    \hat{\mathbf{x}}(c, h, w) = \frac{1}{HW}\sum_{u=0}^{H-1}\sum_{v=0}^{W-1} \hat{\mathbf{X}}(c, u, v)\cdot e^{j2\pi\left(\frac{hu}{H}+\frac{wv}{W}\right)}.
\end{equation}

\subsection{DFT-Based Federated Domain Generalization}\label{sec: DFT-Based Federated Domain Generalization}
\begin{figure}[tb]
\centerline{\includegraphics[width=0.6\linewidth]{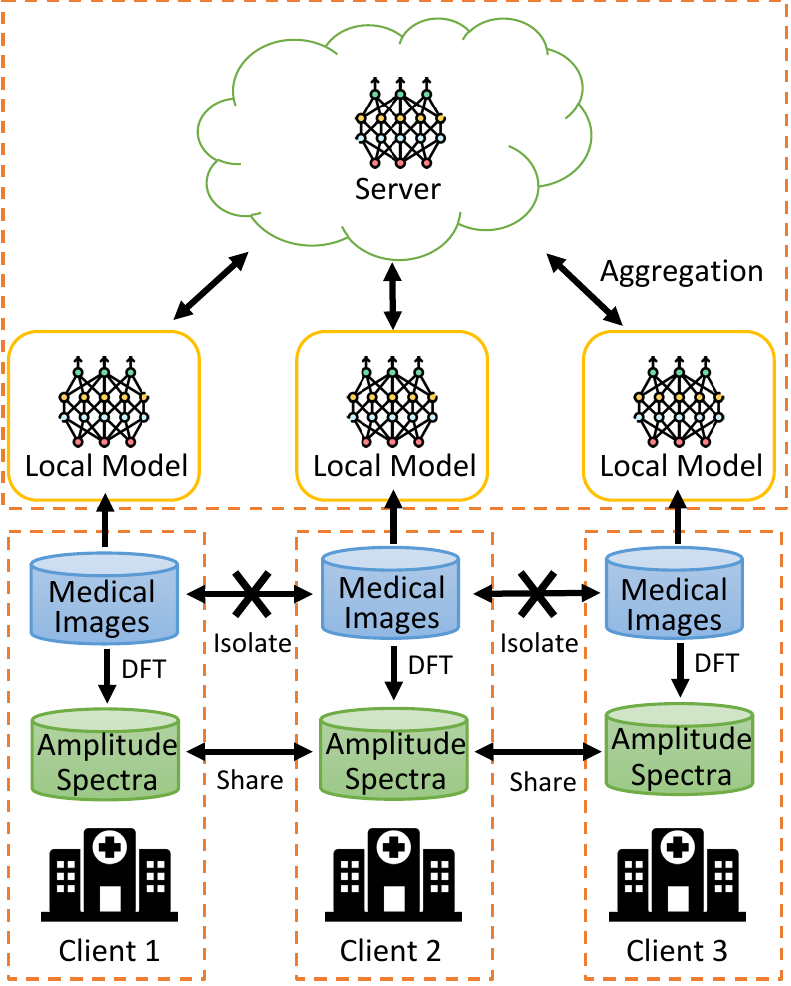}}
\caption{DFT-Based Federated Domain Generalization.}
\label{fig: feddg}
\end{figure}

The FDG approach relies solely on amplitude information from other domains, making it well-suited for federated domain generalization~\cite{liu2021feddg}. The framework is shown in Fig.~\ref{fig: feddg}. This method upholds privacy standards since an image reconstructed using only the amplitude spectrum is incomplete and visually unrecognizable. The phase spectrum, which carries essential spatial positioning details, remains untouched. Additionally, because low-frequency amplitude components hold the most critical information, a binary mask $\mathbf{M}$—with values of 1 for low-frequency regions and 0 elsewhere—can be applied to the target amplitude spectrum. This minimizes the risk of privacy breaches while preserving the most important data for generalization. In another word, Eq. (\ref{eq: fda}) is rewritten as:
\begin{equation}
    \hat{\mathbf{A}} = (1-\lambda)\cdot\mathbf{A}*(\mathbf{1}-\mathbf{M})+\lambda\cdot\tilde{\mathbf{A}}*\mathbf{M}.\label{eq: feddg}
\end{equation}
In this way, only the low-frequency components of the amplitude spectrum are swapped. 

\subsection{Applying Thresholding to Improve DFT-Based FDG}
\begin{figure}[tb]
\centering
\subfloat[Source]{\includegraphics[width=0.19\linewidth]{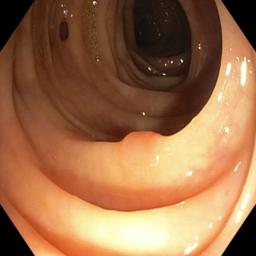}}\vspace{1pt}
\subfloat[Target]{\includegraphics[width=0.19\linewidth]{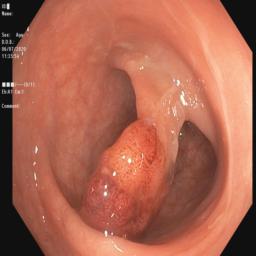}}\vspace{1pt}
\subfloat[DFT]{\includegraphics[width=0.19\linewidth]
{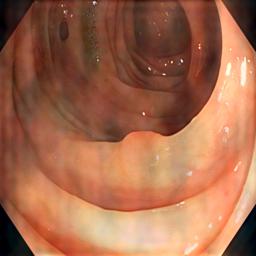}}\vspace{1pt}
\subfloat[DFT+ST]{\includegraphics[width=0.19\linewidth]{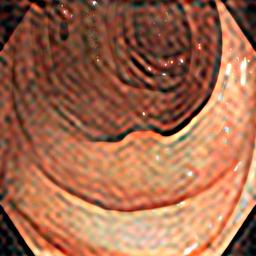}}\vspace{1pt}
\subfloat[DFT+HT]{\includegraphics[width=0.19\linewidth]{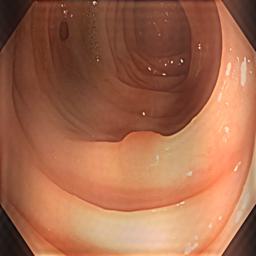}}
\caption{Synthetic images generated by different Fourier-transform-based domain generalization method.}
\label{fig: synthetic images}
\end{figure}

As Fig.~\ref{fig: synthetic images}(c) shows, background inference (shadows) is introduced by the Fourier-Transform-Based domain generalization. In our ICASSP-2024 study~\cite{pan2024domain}, we applied soft-thresholding (ST) filtering to minimize background interference in retinal fundus image segmentation. Soft-thresholding, a widely recognized technique in signal processing and statistical estimation, promotes sparsity and enhances data quality~\cite{donoho1995noising, tomita2022denoising}. By applying a threshold to the DFT coefficients, it effectively reduces noise and irrelevant details while preserving essential features. The soft-thresholding function operates for a given threshold $T$ as follows:
\begin{equation}
\text{SoftThreshold}(x, T) =
\begin{cases}
x - T, & \text{if } x > T \\
0, & \text{if } |x| \leq T \\
x + T, & \text{if } x < -T
\end{cases}
\end{equation}
where $X$ is the input and $T$ is the threshold value. However, when we applied the same domain generalization method~\cite{pan2024domain} on the polyp dataset, we notice that the soft-thresholding function enhances noisy patterns while eliminating the shadows caused by the target background, as Fig.~\ref{fig: synthetic images}(d) present. This is because the soft-thresholding function shrinks all coefficients (both large and small) toward zero. This introduces bias in the retained large coefficients, potentially distorting important features. 

In this work, we replace the soft-thresholding with the hard-thresholding (HT) function~\cite{donoho1996density} to overcome the above issue. A hard threshold is a decision rule in signal processing, machine learning, and data analysis where values are set to zero if they fall below a certain threshold, and remain unchanged otherwise. It's typically used in situations like denoising, feature selection, and sparse signal recovery. In wavelet denoising, hard thresholding is applied to wavelet coefficients by zeroing out all coefficients that have a magnitude less than a set threshold. Formally, for a threshold $T$, the hard-thresholding function is:
\begin{equation}
\text{HardThreshold}(x, T) =
\begin{cases}
x, & \text{if } |x| \geq T \\
0, & \text{if } |x| < T
\end{cases}
\end{equation}
As Fig.~\ref{fig: synthetic images}(d) reveals, applying hard thresholds removes shadows caused by the amplitude swapping without producing noisy patterns, but it may smooth the synthetic images.

\begin{figure}[tb]
\centerline{\includegraphics[width=1\linewidth]{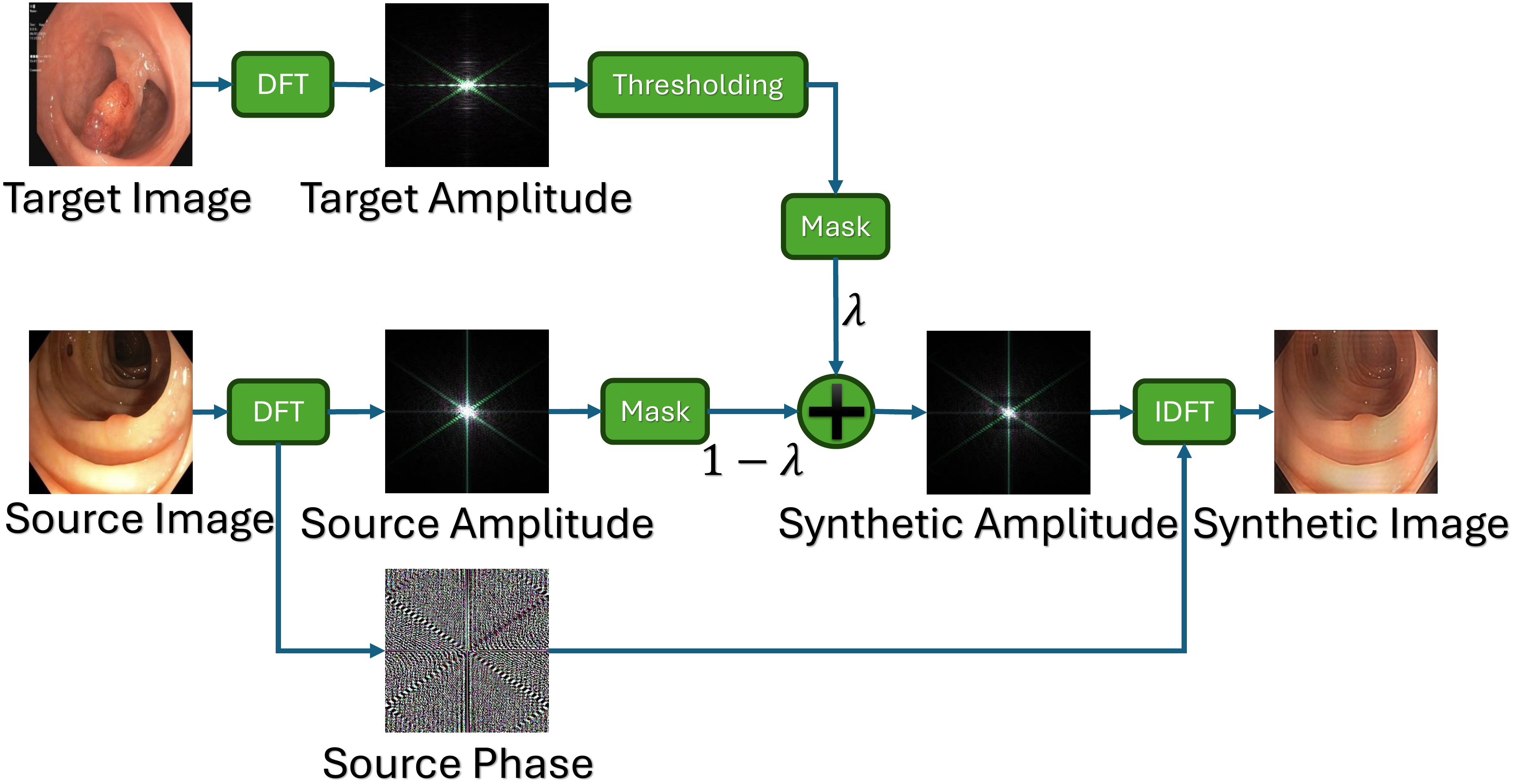}}
\caption{Fourier-Transform-Based Domain Generalization.}
\label{fig: fdg}
\end{figure}

By applying thresholds $\mathbf{T}\in\mathbb{R}^C$ on the target amplitude spectrum, We modify Eq. (\ref{eq: feddg}) as:
\begin{equation}
    \hat{\mathbf{A}} = (1-\lambda)\cdot\mathbf{A}*(\mathbf{1}-\mathbf{M})+\lambda\cdot\text{Threshold}(\tilde{\mathbf{A}}, \mathbf{T})*\mathbf{M}.\label{eq: feddg-st}
\end{equation}
These thresholds $\mathbf{T}$ are determined dynamically based on the largest magnitudes: 
\begin{equation}\label{eq: T=amaxA}
   \mathbf{T}=\alpha\cdot\max(\tilde{\mathbf{A}}),
\end{equation}
where $\alpha\le 5\%$ controls the strength of the thresholding filtering. By integrating the thresholding filtering into the DFT-based FDG, our method is proposed in Algorithm~\ref{al: DFTST} and Fig.~\ref{fig: fdg}.

\begin{algorithm}[tb]
	\caption{Frequency-Based Federated Domain Generalization on $K$ centers.}
	\label{al: DFTST}
	\begin{algorithmic}[1]
		\renewcommand{\algorithmicrequire}{\textbf{Input:}}
		\renewcommand{\algorithmicensure}{\textbf{Output:}}
		
		\REQUIRE Global model $\theta^0$, training datasets $(\Omega_k, \Psi_k)$ for $k=0, 1, \dots, K-1$. Hyperparameters: number of epochs $T$, aggregation weights $a_k$ for $k=0, 1, \dots, K-1$.
		\ENSURE  Well-trained Global model $\theta$.
		\FOR{$t=0, 1, \dots, T-1$} 
            \FOR{$k=0, 1, \dots, K-1$} 
            \STATE Assign weights to local model: $\theta_k^t=\theta^t$;
            \STATE Run Algorithm~\ref{al: FTST} repeatedly to generate $\hat{\Omega}_k$ from $\Omega_k$ and amplitudes from $\{\Omega_j, j=0, 1, \dots, T-1\}$; 
            \STATE Update $\theta_k^t$ using $(\hat{\Omega}_k, \Psi_k)$;
            \STATE Update $\theta_k^t$ using $(\Omega_k, \Psi_k)$;
            \ENDFOR 
             \STATE$\theta^{t+1}=\text{Aggregate}(\theta_0^t, \dots, \theta_{K-1}^t)$:$\theta^{t+1}=\sum_{k=0}^{K-1} a_k\theta_k^t$;
            \ENDFOR \\
	\end{algorithmic}
\end{algorithm}

\begin{algorithm}[tb]
	\caption{Frequency-Based Domain Generalization.}
	\label{al: FTST}
	\begin{algorithmic}[1]
		\renewcommand{\algorithmicrequire}{\textbf{Input:}}
		\renewcommand{\algorithmicensure}{\textbf{Output:}}
		\REQUIRE Source image $\mathbf{x}\in\Omega_k$, arbitrary target amplitude spectrum $\tilde{\mathbf{A}}$ from $\Omega_j$. Hyperparameters: binary mask $\mathbf{M}, \alpha=5\%, \lambda=\text{rand}(0, 1)$.
		\ENSURE  Generated image $\hat{\mathbf{x}}$.
		\STATE $\mathbf{X}=\mathcal{F}(\mathbf{x})$ 
  \STATE $\mathbf{A}=\text{abs}(\mathbf{X})$;
  \STATE $\mathbf{P}=\text{angle}(\mathbf{X})$;
  \STATE $\mathbf{T}=\alpha\cdot\max(\tilde{\mathbf{A}})$;
  \STATE $\hat{\mathbf{A}} = (1-\lambda)\cdot\mathbf{A}*(\mathbf{1}-\mathbf{M})+\lambda\cdot\text{Threshold}(\tilde{\mathbf{A}}, \mathbf{T})*\mathbf{M}$;
  \STATE $\hat{\mathbf{X}}=$  complex$(\hat{\mathbf{A}}, \mathbf{P})$: $\hat{\mathbf{X}}(c, u, v) =  \hat{\mathbf{A}}(c, u, v)\cdot e^{-j\mathbf{P}(c, u, v)}$;
  \STATE $\hat{\mathbf{x}}=\mathcal{F}^{-1}(\hat{\mathbf{X}})$;
	\end{algorithmic}
\end{algorithm}

\section{Experimental Results}
For segmentation backbone, we train TransNetR~\cite{jha2024transnetr} on the PolypGen~\cite{ali2023multi} dataset. The PolypGen dataset contains images from 6 different centers covering varied populations. We split it as 9:1 for training and testing. Furthermore, in order to achieve accurate and real-time polyp segmentation on unseen out-of-distribution data, we further evaluate the model on the Kvasir-SEG~\cite{jha2020kvasir} dataset. FedAvg~\cite{mcmahan2017communication} and FedProx~\cite{li2020federated} were used as FL baselines. We applied different FDG methods in the training for benchmarking: FedDG-ELCFS~\cite{liu2021feddg} applies the Fourier-transform-based domain generalization into federated learning frameworks as Section~\ref{sec: DFT-Based Federated Domain Generalization}. FedDG-GA~\cite{zhang2023federated} applies dynamic aggregation weights. It does not generate synthetic images during the training. Therefore, FedDG-GA can be implemented together with our proposed FedDG methods.

Let $\hat{Y}$ denote the network's output and $Y$ denote the desired output, we evaluate methods using the following metrics: intersection over union (mIoU),  dice similarity coefficient (DSC), recall, precision, F2 score, and Hausdorff distance (HD):

\begin{equation}
\text{IoU} = \frac{|\hat{Y} \cap Y|}{|\hat{Y} \cup Y|},
\end{equation}
\begin{equation}
\text{Recall} = \frac{|\hat{Y} \cap Y|}{|Y|},
\end{equation}
\begin{equation}
\text{Precision} = \frac{|\hat{Y} \cap Y|}{|\hat{Y}|},
\end{equation}
\begin{equation}
\text{DSC} = \frac{2 \cdot \text{Precision} \cdot \text{Recall}}{2 \cdot \text{Precision} + \text{Recall}},
\end{equation}
\begin{equation}
F_2 = \frac{5 \cdot \text{Precision} \cdot \text{Recall}}{4 \cdot \text{Precision} + \text{Recall}},
\end{equation}
\begin{equation}
\text{HD}(\hat{Y}, Y) = \max \left\{ \sup_{a \in \hat{Y}} \inf_{b \in Y} \|a - b\|, \sup_{b \in Y} \inf_{a \in \hat{Y}} \|a - b\| \right\}.
\end{equation}
All these metrics except HD are the higher the better, and HD is the lower the better.

\begin{figure*}[htbp]
\centerline{\includegraphics[width=0.85\linewidth]{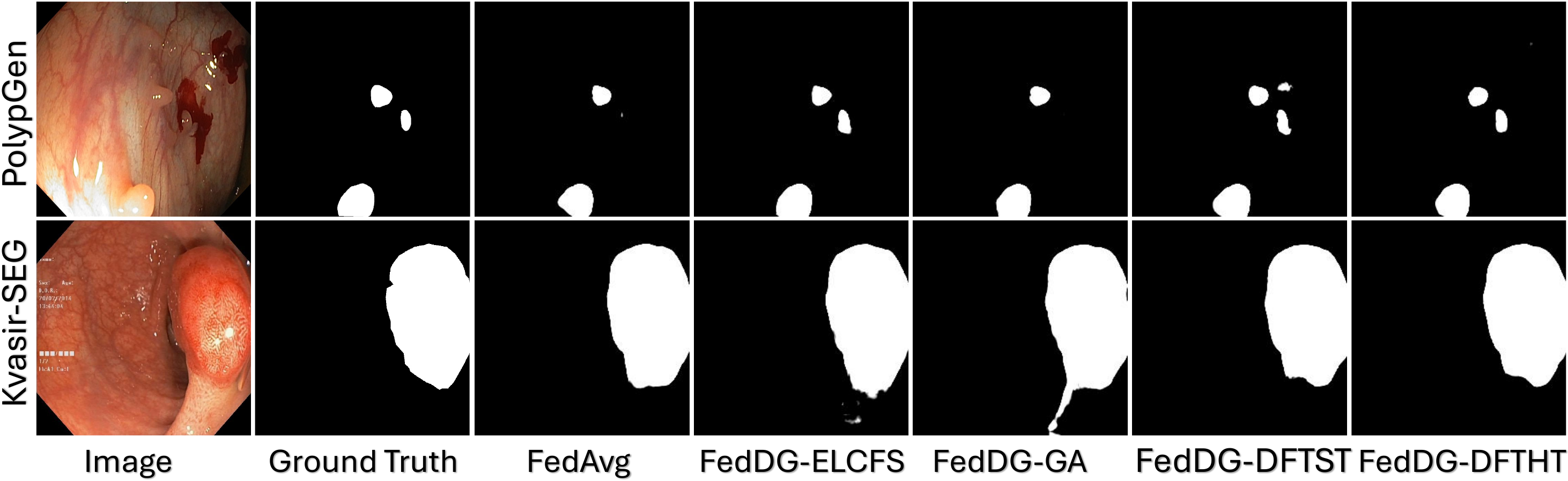}}
\caption{Samples of the polyp segmentation results.}
\label{fig: result}
\end{figure*}

Table~\ref{tab: fed} presents our extensive experiments on the polyp segmentation results under FedAvg and FedProx frameworks and benchmarking results with other algorithms.  For the metrics defined above, we have highlighted the best-obtained results in bold. Fig.~\ref{fig: result} shows samples of the polyp segmentation results. Our comprehensive evaluations reveal that thresholding-based FDG methods (either soft or hard-thresholding) perform better than others. 

\begin{table*}[htbp]
\caption{Polyp segmentation results. The best results are in bold.}
\begin{center}
\begin{tabular}{l|cccccc|cccccc}
\hline
&\multicolumn{6}{c|}{\textbf{Baseline: FedAvg~\cite{mcmahan2017communication}}}&\multicolumn{6}{c}{\textbf{Baseline: FedProx ($\mu=0.3$)~\cite{li2020federated}}}\\
\hline
\textbf{Method}&\textbf{IoU$\uparrow$}  & \textbf{DSC$\uparrow$}  & \textbf{Recall$\uparrow$}& \textbf{Precision$\uparrow$} & \textbf{F2$\uparrow$} & \textbf{HD$\downarrow$}&\textbf{IoU$\uparrow$}  & \textbf{DSC$\uparrow$}  & \textbf{Recall$\uparrow$}& \textbf{Precision$\uparrow$} & \textbf{F2$\uparrow$} & \textbf{HD$\downarrow$} \\
\hline
\multicolumn{13}{l}{\textbf{PolypGen Average}}\\
\hline
Baseline&  0.7797 & 0.8304 & 0.8422 & 0.9094 & 0.8208 & 24.22&\textbf{0.7938} & 0.8452 & \textbf{0.8625} & 0.9078 & \textbf{0.8432} & 23.91\\
+FedDG-ELCFS~\cite{liu2021feddg}&0.7730 & 0.8274 & 0.8513 & 0.9002 & 0.8245 & 24.67&0.7890 & \textbf{0.8459} & 0.8429 & \textbf{0.9209} & 0.8349 & 23.81\\
+FedDG-GA~\cite{zhang2023federated}&\textbf{0.7853} & \textbf{0.8364} & \textbf{0.8522} & 0.8943 & \textbf{0.8299} & 24.30&0.7933 & 0.8453 & 0.8575 & 0.9150 & 0.8420 & \textbf{23.71}\\
+FedDG-DFTST&0.7719 & 0.8246 & 0.8435 & 0.9101 & 0.8189 & 24.81& 0.7811 & 0.8344 & 0.8424 & 0.9160 & 0.8305 & 23.88\\
+FedDG-DFTHT&0.7727 & 0.8226 & 0.8410 & \textbf{0.9114} & 0.8174 & \textbf{23.78} & 0.7672 & 0.8218 & 0.8424 & 0.8870 & 0.8213 & 24.15\\
\hline
\multicolumn{13}{l}{\textbf{PolypGen Center 1, total 256 images, where 230 for train and 26 for test}}\\
\hline
Baseline&0.7576 & 0.8106 & 0.8116 & \textbf{0.9369} & 0.8096 & 33.51 & 0.7559 & 0.7972 & 0.8087 & \textbf{0.9430} & 0.8036 & 33.02 \\
+FedDG-ELCFS~\cite{liu2021feddg}&0.8092 & 0.8604 & 0.8753 & 0.9268 & 0.8687 & 32.91&0.7764 & 0.8310 & 0.8178 & 0.9287 & 0.8224 & 33.36\\
+FedDG-GA~\cite{zhang2023federated}&0.7443 & 0.7897 & 0.7999 & 0.8613 & 0.7952 & 34.65&\textbf{0.8037} & \textbf{0.8566} & \textbf{0.8620} & 0.9337 & \textbf{0.8592} & \textbf{32.66}\\
+FedDG-DFTST&0.7891 & 0.8475 & 0.8624 & 0.9172 & 0.8555 & 35.04&0.7477 & 0.8011 & 0.8046 & 0.9259 & 0.8017 & 34.19\\
+FedDG-DFTHT&\textbf{0.8153} & \textbf{0.8634} & \textbf{0.8768} & 0.9316 & \textbf{0.8706} & \textbf{31.69}&0.7403 &0.7836 & 0.7948 & 0.8583 & 0.7893 & 33.33\\
\hline
\multicolumn{13}{l}{\textbf{PolypGen Center 2, total 301 images, where 270 for train and 31 for test}}\\
\hline
Baseline& 0.8762 & 0.9272 & 0.8929 & 0.9753 & 0.9055 & 21.97&0.8484 & 0.8996 & \textbf{0.9145} & 0.9276 & 0.8881 & 21.96\\
+FedDG-ELCFS~\cite{liu2021feddg}&0.8553 & 0.9098 & 0.8879 & 0.9608 & 0.8944 & 22.25&0.8465 & 0.9051 & 0.8711 & 0.9686 & 0.8831 & 23.31\\
+FedDG-GA~\cite{zhang2023federated}& \textbf{0.8874} & \textbf{0.9367} & \textbf{0.9052} & 0.9767 & \textbf{0.9170} & \textbf{21.16}&\textbf{0.8913} & \textbf{0.9394} & 0.9102 & \textbf{0.9761} & \textbf{0.9212} & \textbf{20.63}\\
+FedDG-DFTST&0.8502 & 0.8987 & 0.8663 & \textbf{0.9791} & 0.8781 & 22.51&0.8541 & 0.9042 & 0.9118 & 0.9365 & 0.8887 & 21.66\\
+FedDG-DFTHT&0.8166 & 0.8654 & 0.8756 & 0.9357 & 0.8509 & 22.90& 0.8684&0.9232 & 0.9098 & 0.9532 & 0.9132 & 22.45\\
\hline
\multicolumn{13}{l}{\textbf{PolypGen Center 3, total 457 images, where 441 for train and 46 for test}}\\
\hline
Baseline&0.8356 & 0.8923 & 0.8877 & 0.9065 & 0.8885 & 23.88&0.8196 & 0.8817 & \textbf{0.9062} & 0.8718 & \textbf{0.8939} & 26.92\\
+FedDG-ELCFS~\cite{liu2021feddg}& 0.8323 & 0.8892 & \textbf{0.8915} & 0.8982 & 0.8896 & 26.05&0.8301 & 0.8902 & 0.8897 & 0.8991 & 0.8889 & 25.86\\
+FedDG-GA~\cite{zhang2023federated}&  \textbf{0.8510} & \textbf{0.9029} & \textbf{0.8915} & 0.9190 & \textbf{0.8955} & 23.96&0.8271 & 0.8884 & 0.8887 & 0.8970 & 0.8875 & 26.17\\
+FedDG-DFTST&0.8316 & 0.8823 & 0.8727 & 0.9171 & 0.8758 & 24.78&\textbf{0.8322} & \textbf{0.8912} & 0.8885 & \textbf{0.9025} & 0.8886 & \textbf{25.18}\\
+FedDG-DFTHT&0.8335 & 0.8833 & 0.8658 & \textbf{0.9271} & 0.8720 & \textbf{23.12}&0.8049&0.8719 & 0.8910 & 0.8735 & 0.8813 & 26.90
\\
\hline
\multicolumn{13}{l}{\textbf{PolypGen Center 4, total 227 images, where 204 for train and 23 for test}}\\
\hline
Baseline& \textbf{0.7440} & \textbf{0.7760} & 0.8970 & 0.8439 & \textbf{0.7695} & \textbf{21.20}&0.7889 & 0.8217 & 0.8982 & 0.8862 & 0.8147 & 18.52\\
+FedDG-ELCFS~\cite{liu2021feddg}& 0.7207 & 0.7566 & 0.8742 & 0.8447 & 0.7476 & 22.84&0.7907 & 0.8240 & \textbf{0.9028} & 0.8842 & 0.8185 & 16.59\\
+FedDG-GA~\cite{zhang2023federated}& 0.6885 & 0.7253 & \textbf{0.9056} & 0.7781& 0.7278 & 22.36&0.7907 & 0.8199 & 0.8993 & 0.8880 & 0.8146 & 17.05\\
+FedDG-DFTST&0.7112 & 0.7522 & 0.8533 & \textbf{0.8546} & 0.7331 & 21.81&\textbf{0.8254} & \textbf{0.8621} & 0.9014 & \textbf{0.9181} & \textbf{0.8588} & \textbf{16.45}\\
+FedDG-DFTHT&0.7096 & 0.7518 & 0.8755 & 0.8250 & 0.7459 & 22.23&0.7360 & 0.7727 & 0.9021 & 0.8284 & 0.7710 & 18.24\\
\hline
\multicolumn{13}{l}{\textbf{PolypGen Center 5, total 208 images, where 187 for train and 21 for test}}\\
\hline
Baseline & 0.5721 & 0.6301 & 0.6139 & 0.8545 & 0.6182 & 26.79&\textbf{0.6225} & \textbf{0.6917} & \textbf{0.6815} & 0.8586 & \textbf{0.6844} & 26.30\\
+FedDG-ELCFS~\cite{liu2021feddg}&0.5437 & 0.6190 & 0.6447 & 0.8302 & 0.6291 & 26.76&0.5776 & 0.6554 & 0.6382 & \textbf{0.8758} & 0.6395 & 25.47\\
+FedDG-GA~\cite{zhang2023federated}& \textbf{0.6023} & \textbf{0.6723} & 0.6533 & 0.8516 & 0.6586 & 27.23&0.5116 & 0.5795 & 0.6328 & 0.8164 & 0.5956 & 28.98\\
+FedDG-DFTST& 0.5978 & 0.6615 & \textbf{0.6976} & 0.8640 & \textbf{0.6788} & 25.96&0.5118 & 0.5734 & 0.5913 & 0.8609 & 0.5815 & 27.61\\
+FedDG-DFTHT&0.5592 & 0.6174 & 0.6350 & \textbf{0.8652} & 0.6257 & \textbf{24.69}&0.5441 & 0.6096 & 0.6087 & 0.8528 & 0.6071 & \textbf{25.10}\\
\hline
\multicolumn{13}{l}{\textbf{PolypGen Center 6, total 88 images, where 79 for train and 9 for test}}\\
\hline
Baseline & 0.7684 & 0.8209 & \textbf{0.9251} & 0.8285 & 0.8167 & 19.33&\textbf{0.8776} & \textbf{0.9310} & 0.9268 & 0.9365 & \textbf{0.9284} & \textbf{15.83}\\
+FedDG-ELCFS~\cite{liu2021feddg}&0.7498 & 0.8071 & 0.9171 & 0.8129 & 0.8058& 17.12&0.8558 & 0.9137 & 0.9091 & 0.9258 & 0.9099 & 16.60\\
+FedDG-GA~\cite{zhang2023federated}&0.8595 & 0.9164 & 0.9238 & 0.9150 & \textbf{0.9199} & 16.95&0.8756 & 0.9295 & 0.9127 & \textbf{0.9476} & 0.9193 & 15.77\\
+FedDG-DFTST& 0.7506 & 0.8046 & 0.9229 & 0.8037 & 0.8080 & 17.05&0.8541 & 0.9133 & 0.9130 & 0.9149 & 0.9130 & 16.88\\
+FedDG-DFTHT&\textbf{0.8721} & \textbf{0.9275} & 0.9117 & \textbf{0.9468} & 0.9176 & \textbf{16.12}&0.8511 & 0.9098 & \textbf{0.9305} & 0.8962 & 0.9211 & 17.39\\
\hline
\multicolumn{13}{l}{\textbf{Kvasir-SEG, total 1000 images, test only}}\\
\hline
Baseline & 0.7882 & 0.8596 & 0.8661 & 0.9083 & 0.8568 & 41.80&0.8007 & 0.8716 & \textbf{0.8871} & 0.9017 & \textbf{0.8728} & 41.03\\
+FedDG-ELCFS~\cite{liu2021feddg}&0.7890 & 0.8618 & \textbf{0.8767} & 0.8993 & \textbf{0.8641} & 42.43&0.7985 & 0.8682 & 0.8793 & 0.9074 & 0.8672 & 41.20\\
+FedDG-GA~\cite{zhang2023federated}&0.7888 & 0.8621 & \textbf{0.8767} & 0.8977 & 0.8633 & 42.56&\textbf{0.8024} & \textbf{0.8730} & 0.8702 & \textbf{0.9185} & 0.8663 & \textbf{40.28}\\
+FedDG-DFTST& 0.7876 & 0.8611 & 0.8688 & 0.9035 & 0.8594 & 42.47&0.8006 & 0.8713 & 0.8724 & 0.9156 & 0.8666 & 40.38\\
+FedDG-DFTHT&\textbf{0.7950} & \textbf{0.8658} & 0.8689 & \textbf{0.9116} & 0.8621 & \textbf{41.27}&0.7998 & 0.8702 & 0.8856 & 0.9009 & \textbf{0.8728} & 40.69\\
\hline
\end{tabular}
\label{tab: fed}
\end{center}
\end{table*}

\section{Conclusion}
In this paper, we present a frequency-based Federated Domain Generalization (FDG) approach for polyp segmentation, utilizing soft- and hard-thresholding in the Fourier domain to reduce background noise and strengthen model robustness. By preserving essential frequency components and filtering out irrelevant background signals, our method effectively addresses domain shifts. Experimental results show marked improvements in segmentation accuracy and generalization across diverse medical imaging datasets compared to traditional methods. This study underscores the promise of incorporating frequency domain techniques into federated learning, offering a more adaptable and resilient framework for medical image analysis.

\bibliographystyle{unsrt}
\bibliography{ref}

\end{document}